\documentclass[letterpaper]{article} 
\usepackage{aaai25}  
\usepackage{times}  
\usepackage{helvet}  
\usepackage{courier}  
\usepackage[hyphens]{url}  
\usepackage{graphicx} 
\usepackage{natbib}  
\usepackage{caption} 
\usepackage{algorithm}
\usepackage{algorithmic}
\usepackage{amsmath}
\usepackage{amsthm}
\usepackage{amsfonts}
\usepackage{color}
\usepackage{multirow}
\usepackage{makecell}
\def\BibTeX{{\rm B\kern-.05em{\sc i\kern-.025em b}\kern-.08em
    T\kern-.1667em\lower.7ex\hbox{E}\kern-.125emX}}

\usepackage{newfloat}
\usepackage{listings}
\DeclareCaptionStyle{ruled}{labelfont=normalfont,labelsep=colon,strut=off} 
\floatstyle{ruled}
\newfloat{listing}{tb}{lst}{}
\floatname{listing}{Listing}
\title{SoMeR: A Multi-View Social Media User Representation Learning Framework}
\author{
    Siyi Guo\textsuperscript{\rm 1},
    Keith Burghardt\textsuperscript{\rm 1},
    Valeria Pantè\textsuperscript{\rm 1},
    Kristina Lerman\textsuperscript{\rm 1}
}
\affiliations{
    \textsuperscript{\rm 1}USC Information Sciences Institute\\

    Marina del Rey, California, USA\\
    siyiguo@isi.edu
}

\usepackage{bibentry}

\begin{document}

\maketitle

\begin{abstract}
Social media user representation learning aims to capture user preferences, interests, and behaviors in low-dimensional vector representations. These representations are critical to a range of social problems, including predicting user behaviors and detecting inauthentic accounts. However, existing methods are either designed for commercial applications, or rely on specific features like text contents, activity patterns, or platform metadata, failing to holistically model user behavior across different modalities. To address these limitations, we propose SoMeR, a Social Media user Representation learning framework that incorporates temporal activities, text contents, profile information, and network interactions to learn comprehensive user portraits. SoMeR encodes user post streams as sequences of time-stamped textual features, uses transformers to embed this along with profile data, and jointly trains with link prediction and contrastive learning objectives to capture user similarity. We demonstrate SoMeR's versatility through three applications: 1) Identifying information operation driver accounts, 2) Measuring online polarization after major events, and 3) Predicting future user participation in Reddit hate communities. SoMeR provides new solutions to better understand user behavior in the socio-political domains, enabling more informed decisions and interventions.
\end{abstract}

%

\section{Introduction}
Understanding and modeling user behavior on social media is a critical challenge for platforms, researchers, and society at large. As social media continues to shape public discourse, civic processes, and responses to disasters, the ability to accurately represent user preferences, opinions, and behavioral patterns has become increasingly urgent. These user representations serve as foundational building blocks for numerous downstream applications, from personalized content recommendation and detecting social bots~\cite{nwala2023language} to identifying suicidal ideation~\cite{sawhney-etal-2021-suicide} and hate speech~\cite{qian2018leveraging,del2019you}. The complexity of this challenge stems from the multi-faceted nature of user behavior---individuals express themselves through various modalities including text posts, sharing patterns, temporal rhythms of activity, and social connections, all of which evolve dynamically over time.

To address this problem, researchers have developed methods for user representation learning (URL), which aim to learn low-dimensional vector representations that capture user preferences, interests, and behaviors~\cite{li2021survey}. These representations are popular in commercial applications, such as personalized recommendations and targeted advertising~\cite{bhargava2015and,wang2015collaborative,hou2022towards,shin2023scaling}. Applications of URL extend well beyond the commercial sector to the social and political domains, where they are used to uncover latent factors of online behavior that give insights into public attitudes and societal trends
~\cite{pan2019social}. 

However, most advanced URL approaches, designed for recommendation and built on commercially relevant features~\cite{li2021survey,purificato2024user}, are not well-suited for social media. In addition, the current URL methods developed for social media usually depend on a subset of user features, such as text~\cite{hallac2021user}, images~\cite{pan2019social}, activity~\cite{nwala2023language}, or platform-specific features~\cite{almahmoud2018tsim,dahiya2022contextual}. Multi-view approaches exist but are restricted to combining only text and network features, and many remain task- or platform-specific~\cite{feng2021satar,shen2023uniskgrep}. As a result, there is a scarcity of generalized approaches for socio-political problems that (1) holistically model user behavior by combining multiple features from content, temporal activity and user interactions, and (2) can be flexibly applied to different research scenarios.

Creating universal multi-view user representation learning that combines multiple features poses a number of challenges. The social media user populations are highly heterogeneous, composed of individuals with different beliefs, attitudes, interactions, and behaviors. A few users within a population are prolific posters while the vast majority are not. As a result, the temporal user activity is sparse, which presents a challenge when incorporating activity features into a representation. Moreover, the number of connections varies dramatically between users and also changes from time to time, which poses challenges to incorporating network information. Finally, ground truth data to train classifiers is often difficult to obtain, which makes a supervised model representation challenging.

To address these challenges, we propose \textbf{SoMeR}, \textbf{So}cial \textbf{Me}dia user \textbf{R}epresentation learning framework, which incorporates 1) temporal activities, 2) post text, 3) profile information and 4) network interactions to learn a comprehensive portrait of online users. These features are universal across different social platforms, making our framework flexible and adaptable. This framework allows us to discover similar users in populations with heterogeneous beliefs, attitudes, and behaviors.

The method works as follows. We first encode user posts as a sequence of triplets of the form \textit{(timestamp, textual feature, value)}, which augments typically limited time series data by incorporating a variety of features from each post. 
We encode the contextual information of these triplets into an embedding using a transformer-based architecture~\cite{vaswani2017attention}. We combine this triplet embedding with a user profile embedding, and impose two jointly trained objectives: (1) network link prediction to learn interactions between users, and (2) contrastive learning to pull similar users closer and push dissimilar users farther away. As a result, our model learns an embedding space that is aware of user similarity and heterogeneity across temporal, textual, network connection, and user profile dimensions. This pre-training step can be used in unsupervised settings where annotated data is hard to obtain. Moreover, by fine-tuning our model on a few supervision data samples, we can adapt it to various downstream tasks.

We demonstrate the generalizability and versatility of the framework by studying three novel and diverse research questions on two different social media platforms. 
We first identify Information Operation (IO) drivers, which are malicious accounts used to amplify societal divisions or sway public opinion~\cite{wen2020time}. We train a SoMeR-based model on a novel ground truth information operations dataset and detect IO drivers with accuracy that matches or exceeds baselines. 
We next use SoMeR to measure how events deepen polarization. We analyze discussions on X about the US Supreme Court (SCOTUS) ruling overturning federal abortion protections, and find that the ruling increased partisan differences in text and activity of users. 
Finally, we predict future user participation in Reddit hate communities (subreddits). Prior research shows a direct link between online hate and offline attacks \cite{muller2021fanning}, thus detecting who will become active in hate communities can improve interventions to reduce offline hate. SoMeR achieves unexpectedly high accuracy, even if users do not post months before posting their first hate group.

Our framework advances the state-of-the-art by filling gaps in generalizability within URL research and capability in the socio-political domains:
\begin{enumerate}
    \item Our framework integrates multi-view features into a single embedding, enabling the model to include in analysis user opinions and emotions, dynamic activity patterns, demographic profiles, and network interactions, making it better suited for tackling socio-political questions compared to existing models.
    \item We cohesively integrate advanced methodologies into a unified framework, learning temporal behaviors even from sparse activity data using triplet representations, employing transformer architectures for embedding, and utilizing contrastive learning and network link prediction to capture user similarity more effectively.
    \item Unlike task-specific approaches, our framework is universal and adaptable to a broad range of applications.
    \item Our framework has demonstrated scalability, handling datasets with up to 17 million texts.
\end{enumerate}

Our proposed framework helps to bridge the divide between user representation learning and socio-political analysis by leveraging features from dynamic and heterogenous users to improve classification. This work paves the way for more informed decisions and interventions. 
The source code is available at \url{https://anonymous.4open.science/r/ts_clustering-1C10}.


\section{Related Work}\label{section:related_works}

User representation learning has gained widespread interest within the recommender systems due to its ability to capture meaningful and compact embeddings of users' behaviors and preferences~\cite{yuan2021one}. Over the years, researchers have developed many methods, mostly for commercial applications~\cite{li2021survey,purificato2024user}. These include matrix and tensor factorization~\cite{bhargava2015and}, 
auto-encoders~\cite{zhuang2017representation}, 
transformer-based architectures~\cite{cheng2021transferable,hou2022towards,shin2023scaling}, contrastive learning~\cite{oord2018representation,cheng2021transferable}, graph neural networks~\cite{liu2023enhancing}, and large language models (LLM)~\cite{ren2024representation}. Improving from task-specific methods~\cite{guo2017deepfm}, 
researchers have also explored universal user representation learning methods which can be generalized to different downstream tasks in recommendations~\cite{yuan2021one,hou2022towards,shin2023scaling,fazelnia2024generalized}. 

Beyond recommendations, URL also presents considerable opportunities in the social and political domains, enabling a deeper understanding of public sentiment and societal trends~\cite{pan2019social}. However, the approaches designed for recommendations are built on commercially relevant features and/or pre-trained using commercial data~\cite{li2021survey,purificato2024user}, and thus are not compatible in social analyses. Moreover, most current approaches in social domains rely on a narrow subset of user features. 
For example, \citet{mueen2016awarp} and \citet{nwala2023language} uses temporal activity features to detect online bot and IO drivers; \citet{hallac2021user} 
experimented with different textual embedding methods; 
\citet{perozzi2014deepwalk} and \citet{wang2017community} uses network features and node prediction to learn user interests or communities. 
While multi-view studies are available, they are typically limited to integrating only text and network features, with many being tailored to specific tasks or platforms~\cite{ribeiro2018characterizing,wang2019online,lai2020merl,feng2021satar,shen2023uniskgrep}. 
There clearly lacks a universal framework for social analysis that is generalizable to different platforms and downstream tasks, as well as one that incorporates all of the textual, temporal, profile and network features into a multi-view embedding. Our framework, \textbf{SoMeR}, addresses these gaps.



\section{Methods} \label{section:methods}
\begin{figure*}[ht]
\centering
\includegraphics[width=1\linewidth]{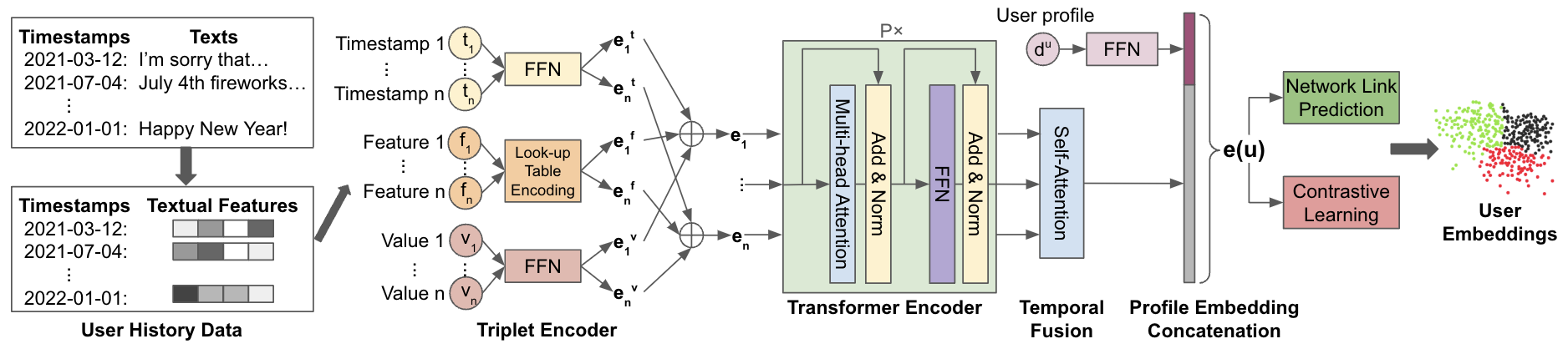}
\caption{Model Architecture of \textbf{SoMeR}. We format a user's posting history into triplets of time, feature, and value, which undergo encoding via a Triplet Encoder, a transformer-based contextual learning module and a fusion attention layer, becoming a user history embedding that is then concatenated to the user profile embedding. Through training with two self-supervised objectives - network link prediction and contrastive loss - our method effectively captures user similarity in the latent space.}
\label{fig:model_arch}
\end{figure*}

We propose a self-supervised framework to learn a latent user embedding space based on the architecture shown in Figure~\ref{fig:model_arch}. From each user's history, a timeline of texts, we first extract certain textual features, such as the sentence embeddings. Next, to better learn from users with sparse activities, we format the textual features and timestamps into  triplets of observations \((timestamp, feature, value)\). These triplets pass through a Triplet Encoder, a transformer-based contextual learning module, and a fusion attention layer, being encoded into a user history embedding. We concatenate it with user's profile embedding from a separate module, obtaining a complete user embedding. We train these encoding modules with two self-supervised objectives: \textit{network link prediction} that learns patterns of interactions, including sharing, following or other connections, and \textit{contrastive loss} that learns user similarity with respect to posting history. This method learns user similarity in a heterogeneous user population without the need for time-consuming human annotations. The method can be easily adapted for various downstream tasks such as supervised classification and unsupervised similarity search.

\subsection{User History Data Processing}
In this study, we consider each user's history to be a collection of timestamped texts. Each text can be an original post, a repost, a reply, etc. There can be other types of user activity, such as likes or views, that also provide valuable information that can be incorporated in future works. To learn from a user's history, we first extract textual features that are useful for a given downstream task. 
In our experiments, we find that using contextual sentence-BERT embeddings \cite{reimers2019sentence} leads to powerful representation. However, BERT embeddings yield complex high-dimensional features, which can slow performance. To reduce model size and complexity, we perform dimensionality reduction using principal component analysis (PCA) to reduce BERT embeddings to the first five components and treat them as textual features. Prior works~\cite{grootendorst2022bertopic} have shown that meaningful contextual representation is still retained after dimension reduction operations on BERT embeddings.

\subsection{Triplet Data Encoder}
Social media users exhibit highly diverse posting behaviors. A small fraction of users generate most of the contents, whereas majority of users only have few posts over a long period of time \cite{krishnan2018insights}. This leads to a sparse matrix of multivariate time series as the input data, making it hard to accurately learn an embedding space from sparse signals, and posing a challenge in time complexity. We take inspiration from \cite{tipirneni2022self} and represent user posting history as a collection of triplets \((timestamp, feature, value)\). This only takes the observations available in the data, allowing us to account for the times with little to no activity. Thus, a dataset of $M$ users can be represented as $U$:
\begin{eqnarray}\label{eq:user_rep}
\mathbf{U}=\{(d^u,\mathbf{X}^u)\}_{u=1}^{M} \ where \ \mathbf{X}=\{(t_n,f_n,v_n)\}_{n=1}^{N}
\end{eqnarray}
Each user $u$ is characterized by their profile feature vector $d^u \in \mathbb{R}^D$ and their posting history $\mathbf{X}^u$, which is a set of $N$ triplets including the timestamp $t_n$, the feature category $f_n$ and the value of this feature category $v_n$. There can be more than one feature-value pairs at one time point, and therefore $N$ may be greater than the total number of time points.

\citet{tipirneni2022self} shows the effectiveness of using a feed-forward network to embed continuous values. We therefore use two separate feed-forward networks to encode timestamp $t_n$ and value $v_n$ into time embedding $e^t_n \in \mathbb{R}^K$ and value embedding $e^v_n \in \mathbb{R}^K$ respectively, where $K$ is the hidden dimension of these embeddings. These networks have one linear layer followed by a tanh activation function. For the feature categories, we use a lookup-table encoder to generate a feature embedding $e^f_n \in \mathbb{R}^K$. Lastly, we add up these three embeddings to be the triplet embedding $e_n^{triplet} \in \mathbb{R}^K$. Adding them up, rather than concatenating, has been proven its effectiveness~\cite{tipirneni2022self}, and allows us to reduce dimensions of model weights to save training time and memory.
\begin{eqnarray}\label{eq:triplet_embedding}
e^t_n & = & W_1^t \ tanh(W_2^tt_n+b^t) \\
e^v_n & = & W_1^v \ tanh(W_2^vv_n+b^v) \\
e^f_n & = & LookupEncoder(f_n) \\
e_n^{triplet} & = & e^t_n + e^v_n + e^f_n
\end{eqnarray}

\subsection{Transformer Encoder}
The transformer architecture has been shown to have great performance in representation learning for time series and user behavior sequences~\cite{zerveas2021transformer,shin2023scaling}. Therefore we choose to use the transformer encoder to better extract a latent representation from the triplet embedding $e_n^{triplet} \in \mathbb{R}^K$. We use $L$ transformer layers. Each layer has $H$ attention heads with learnable key, query and value weights $W^k,W^q,W^v \in \mathbb{R}^{K \times P}$ where $P$ is the hidden dimension size for these weights. After all attention heads are added up and layer normalized, the embedding vector is projected back to $K$-dimension through a feed-forward network. This network includes two layers with hidden dimension $2K$ and a ReLU activation in the middle. Lastly, layer normalization is applied again. From this transformer module we obtain $e_n^{trans} \in \mathbb{R}^K$. In our experiments and applications, data sizes are within 20 million, and therefore we choose to have a small transformer module with $L=2$, $H=4$ and $P=K/H$. We choose the hidden dimension $K=64$ with a grid search in $[32, 64, 128]$. We show in three applications that these parameters work well for our purposes. 

\subsection{Temporal Fusion}
After all triplets for a user pass through the triplet encoder and the transformer encoder, we use an attention layer to learn the correlations between different triplets. This gives us the integrated embedding of posting history $e^{hist}  \in \mathbb{R}^{K}$ for each user.
\begin{eqnarray}\label{eq:attn_fusion}
a_n & = & W_1^{attn} \ tanh(W_2^{attn}e_n^{trans}+b^{attn}) \\
a_n & = & \frac{exp(a_n)}{\sum_{i=1}^{N}exp(a_i)} \\
e^{hist} & = & \sum_{n=1}^{N}a_n \ e_n^{trans}
\end{eqnarray}
where $W_1^{attn} \in \mathbb{R}^{2K}$, $W_2^{attn} \in \mathbb{R}^{2K \times K}$, $b^{attn} \in \mathbb{R}^{2K}$ are trainable weights and intercept.

\subsection{Profile Embedding}
Other than the posting history of a user, their profile features, e.g., location and number of followers and friends, can also play an important role. Therefore, we add a feed-forward network to learn a profile embedding $e^{prof} \in \mathbb{R}^{K}$ from user's profile vector $d^u \in \mathbb{R}^{D}$, and concatenate it to the user history embedding $e^{hist}$ to obtain a complete user embedding $e^u  \in \mathbb{R}^{2K}$.
\begin{eqnarray}\label{eq:prof_embedding1}
e^{prof} & = & W_1^{prof} \ tanh(W_2^{prof}d^u+b^{prof}) \\
e^u & = & e^{hist} \oplus e^{prof}
\end{eqnarray}
where $W_1^{prof} \in \mathbb{R}^{2K}$, $W_2^{prof} \in \mathbb{R}^{2K \times D}$, $b^{prof} \in \mathbb{R}^{2K}$ are trainable weights and intercept, and $\oplus$ is a concatenation operation.

\subsection{Network Link Prediction}
Individuals connected in social networks tend to be similar, and their behaviors are often affected by other users in the network~\cite{mcpherson2001birds}. Therefore it's crucial to include network connection features when learning user representations~\cite{almahmoud2018tsim}. We design a self-supervised network link prediction objective to train our model to learn interaction activities such as sharing, following and commenting. It is a feed-forward module to perform link prediction, with a binary cross-entropy loss. During training, we consider all pairs of distinct users in each batch. The feature for link prediction is the concatenated embedding of a user pair, and we obtain the links from a self-defined network as the binary labels (e.g. whether a user repost another user). This is described as following:
\begin{align}
& \tilde{y}^{link}_{i,j} = \sigma(W_1^{link} \ ReLU(W_2^{link}(e^u_i \oplus e^u_j)+b^{link}))
\end{align}
\begin{equation}
\begin{split}
\mathcal{L}_{\text{network}} & = - \frac{1}{N_{\text{distinct pairs}}} \sum_{\substack{i,j \in batch\\i \neq j}} \biggl[ y^{link}_{i,j} \cdot log(\tilde{y}^{link}_{i,j}) \\
& + (1-y^{link}_{i,j}) \cdot log(1-\tilde{y}^{link}_{i,j}) \biggl]
\end{split}
\end{equation}
where $i,j$ are indices of two users in a training batch, $W_1^{link} \in \mathbb{R}^{2K}$, $W_2^{link} \in \mathbb{R}^{2K \times 4K}$ and $b^{link} \in \mathbb{R}^{2K}$ are trainable weights and intercept, $\sigma(\cdot)$ is the sigmoid function, $N_{\text{distinct pairs}}$ is the number of all pairs of distinct users in a batch, and $y^{link}$ is the binary label of whether $i$ and $j$ are connected in the network.

\subsection{Contrastive Learning}
Contrastive learning aims to obtain a latent embedding space in which similar samples are closer and distinct samples are farther from each other. Many prior works in user representation learning have shown its success~\cite{cheng2021transferable,shin2023scaling}. We adopt self-supervised contrastive learning to learn user similarity in their posting histories. The InfoNCE \cite{oord2018representation} loss function uses categorical cross-entropy loss to optimize the negative log probability of classifying one positive or similar sample correctly among a set of negative or unrelated samples. In our case, it can be written as:
\begin{equation}\label{eq:prof_embedding2}
\mathcal{L}_{\text{InfoNCE}} = - \frac{1}{\text{batch size}} \sum_{i \in \text{batch}} \left[log \frac{exp(e^{u}_ie^{u^+}_i/\tau)}{\sum_{j \in batch, i \neq j}exp(e^u_ie^u_j/\tau)}\right]
\end{equation}
For $e^u_i$ a user history embedding, $e^{u^+}_i$ is its positive pair. $e^u_j \ \forall j \in batch\ \text{where}\ i \neq j$ are the embeddings of all other users in that randomly retrieved batch, which we consider as negative samples to $e^u_i$. We further perform temperature scaling with parameter $\tau$, which is tuned during training using grid search in $[0.5, 1, 3, 6]$.

To generate the positive sample paired to each user history embedding, we perform data augmentation on users' triplet data. Although researchers have proposed many time series data augmentation methods~\cite{wen2020time}, many classical methods are not applicable to our scenario. For example, a user history cannot be sub-sequenced or shuffled in time domain. We use another efficient and simple way --- bootstrapping with replacement from existing triplets $\{(t_n,f_n,v_n)\}_{n=1}^{N}$. This ensures the generated set of triplets is similar. We incorporate noise into augmentation by 
\begin{enumerate}
    \item varying the number of random draws between $range(start=(1-\gamma)N, \ end=(1+\gamma)N, \ step=1)$, where $N$ is the total number of triplets for a user and $\gamma$ is a hyperparameter to control the noise level,
    \item scaling the value $v_n$ by a factor randomly selected between $range(start=1-\gamma, \ end=1+\gamma, \ step=0.5)$, and
    \item imposing a lag time, randomly selected between 1 -3 days, on the timestamp $t_n$ of sampled triplets.
\end{enumerate}
During training, we tune $\gamma$ by grid searching in $[0.1,0.5,1,2,5]$. The results of the three applications below shows the effectiveness of this augmentation method. 

\subsection{Model Training} \label{section:model_training}
Finally, the contrastive objective function and the network link prediction objective are jointly trained at the same time. The overall loss is
\begin{eqnarray}\label{eq:prof_embedding3}
\mathcal{L} = \mathcal{L}_{\text{InfoNCE}} + \lambda \mathcal{L}_{Network}
\end{eqnarray}
where $\lambda$ is a hyperparameter to balance between two losses. We perform grid search on the IO driver dataset for $\lambda$ in $[0.1, 1, 5, 10]$ and find $\lambda=1$ gives the lowest validation loss. We use Adam optimizer, a learning rate of 5e-5, a cosine decay learning scheduler, an early stopping mechanism monitored by overall loss function, a maximum epoch of 60, and a batch size of 128. See Appendix Table~\ref{tab:hyperparameters} for a summary of all hyperparameters and their selecting procedures. We find changes to these hyperparamters do not strongly affect results. 

\subsection{Model Validity}
To verify that our self-supervised framework indeed learns \emph{both} temporal activity and textual features from the triplet data, we test it on four synthetic datasets. We use synthesized numerical features to mimic textual features in real data. The four datasets simulate (1) a simple scenario with three clusters that are easier to distinguish, (2) a hard scenario with 10 clusters and more noise, (3) a scenario where clusters vary in temporal activity pattern but have the same features and values, and (4) a scenario where clusters vary only in features and values, but having all the same temporal patterns. On all datasets, our model can successfully detect the clusters, indicating it is able to learn heterogeneity in both temporal activities and numerical feature values (see Appendix). 

In the next sections, we use three applications - IO driver detection, political polarization analysis, and hate group joiner prediction to illustrate the effectiveness and versatility of our framework.


\section{Detecting Information Operation (IO) Drivers} \label{section:coordination}
Information operations use strategically organized efforts to manipulate public opinions at scale, contaminating the online information ecosystem with disinformation. The rapidly expanding social media platforms provide fertile grounds for these operations~\cite{pacheco2021uncovering}, necessitating their identification. IO drivers use different tactics from each other and from authentic accounts thus leading to distinct behaviors flagged by previous researchers, such as (1) co-sharing the same posts or URLs, (2) using an identical sequence of hashtags, (3) synchronized behaviors, and (4) sharing very similar posts
~\cite{pacheco2021uncovering,nwala2023language,luceri2023unmasking}. 
Our framework captures temporal, textual, profile, and network features, identifying platform-agnostic synchronized behaviors. We apply supervised fine-tuning on this framework and test it across three campaigns on X to demonstrate its effectiveness.

\subsection{Data}
We evaluate our method on an X dataset that prior IO driver detection methods have benchmarked on~\cite{nwala2023language,luceri2023unmasking}. These drivers of multiple information operations in 21 countries were suspended and released by X because they were associated with malicious IOs and violated the platform terms. We select four campaigns --- one based in China involving a large number of accounts, one small operation in Egypt and UAE, and two operations in Venezuela which are combined into one dataset for testing the multi-campaign scenario. 

We collect a set of control users by first collecting the top five keywords and top five hashtags for each IO driver within our X datasets. We then extract 10 random posts that were posted within the timeframe that the IO driver was active (between their first and last post). For each post, we find the post author and query all their posts made within the timeframe the IO driver is active. Table~\ref{tab:coord_data} shows the information about these campaigns. We randomly split data into 70\% training, 15\% validation and 15\% testing.

\begin{table}[ht]
    \centering
    \caption{Meta-data of Information Operation Datasets}
    \resizebox{0.95\linewidth}{!}{
    \renewcommand{\arraystretch}{1.2}
    \begin{tabular}{ccccc}
        \Xhline{1pt}
        \textbf{Campaign} & \textbf{Time Range} & \textbf{\# IO Drivers} & \textbf{\# Control Users} & \textbf{\# Posts} \\ \hline
        China & 2019 - 2021 & 2016 & 11366 & 17M \\ 
        Egypt-UAE & 2016 - 2019 & 240 & 2164 & 4.5M \\ 
        Venezuela & 2017 - 2021 & 275 & 4183 & 10M \\ \Xhline{1pt}
    \end{tabular}}
    
    \label{tab:coord_data}
\end{table}

\subsection{Pre-training and Supervised Classification}\label{section:coord_methods} We use a two-step approach to classify IO drivers. First, we pre-train our model in the self-supervised manner with both IO driver and control users using all of their posts and meta-data, but without any IO label. For textual post preprocessing, we remove user mentions, URLs, emojis and all non-ASCII characters, but retain the hashtags. This is a multi-lingual datasets including more than 50 languages. Therefore we compute the sentence-BERT embedding for each post using the \texttt{stsb-xlm-r-multilingual} \footnote{\url{https://huggingface.co/sentence-transformers/stsb-xlm-r-multilingual}}, and compute the first five components from PCA as its textual features. Due to GPU limit, we use this technique to reduce the number of triplets and computation complexity. We aggregate data for each user by summing up their PCA embeddings of their posts in 3-day intervals and taking the middle day in the interval as the corresponding timestamp. We have conducted experiments with 1-day, 3-day and 5-day window sizes, and choose to use 3-day as it keeps the data relatively fine-grained and also aggregate data well for low-activity accounts. This gives us the temporal and textual features. For profile features, we use number of followings and followers, as accounts such as news media outlets with lots of followers can behave very differently from other types users. For network link prediction part, we use the repost network (i.e. retweet network) because co-repost has been shown as a potential indicator of an IO driver~\cite{pacheco2021uncovering}. With all these features, we train the model to learn a user embeddings space in which users with similar behaviors are closer.

In the next step, we perform supervised fine-tuning on the learned model parameters with an additional two-layer feed-forward network for binary classification. This network has a linear layers with a hidden dimension of 128, a ReLU activation, a dropout layer with rate of 0.3, a batch normalization, and a second linear layer that project embeddings onto $\mathbb{R}$ for binary prediction. The sigmoid function is then applied and a binary cross-entropy function is used as the loss. We use same hyperparameters described in Methods.

\subsection{Model Performance and Ablation}

\paragraph{Baseline Models}
We compare our method with common IO driver detection methods~\cite{pacheco2021uncovering}. Based on the observation that IO drivers have abnormally similar behaviors, these methods identify users with unusual similarity based on (1) co-repost and (2) co-URL sharing behaviors if cosine similarity is above or at the 99.5-th percentile, (3) hashtag sequence (using a minimum sequence of 5 identical hashtags in the same order within a post) and (4) text similarity in averaged BERT embeddings over all posts from a user, with a cosine similarity threshold of 0.7 (parameters based on prior works \cite{pacheco2021uncovering,luceri2023unmasking}.)

In addition, we also compare with a strong multi-view URL baseline SATAR~\cite{feng2021satar}. This approach combines textual, profile and network features, and the authors uses it for social bot detection. However, the major differences between SATAR and our framework are (1) SATAR does not utilize temporal features; (2) SATAR uses follower count prediction as the self-supervised objective, which is suitable for bot detection but less generalizable for other use cases, whereas SoMer uses both contrastive learning and network prediction objectives for self-supervision signals; (2) SATAR uses the Long Short-Term Memory model (LSTM)~\cite{graves2012long} with word2vec embeddings~\cite{church2017word2vec} whereas SoMer uses more advanced transformer module to learn contextual information from user posting histories. The details of evaluating SATAR are included in Appendix.

\subsubsection{Performance}
Table~\ref{tab:coord_performance} shows the performance of different methods. Our models are evaluated on 10 random training-validation-test data splits. First we compare our method with the baselines. For China and Venezuela campaigns, we outperform or match all the baselines. 
For the gap between the text similarity F1-score and ours for the Egypt-UAE campaign, we hypothesize that it is due to our method utilizing only the first five PCA components on post BERT embeddings, versus the text similarity baseline using the full BERT embeddings. In addition, Egypt-UAE is the smallest campaign, and pre-training on a small dataset can also result in less ideal performance (see more analysis in Appendix). 
Moreover, our full model outperforms SATAR in all scenarios. We observe that SATAR's pre-training step is ineffective, with the fine-tuning step contributing most significantly to its overall performance. SATAR's self-supervised pre-training objective, which predicts user follower counts, easily leads to overfitting. For example, for the Chinese campaign, F1-score for predicting follower counts reaches 0.88 but the F1 for predicting IO drivers is only 0.13 during pre-training. After fine-tuning, the F1 for predicting IO drivers reaches 0.98 as in Table~\ref{tab:coord_performance}. This reliance on labeled data highlights SATAR’s limited generalizability to scenarios with sparse labels. In conclusion, the consistently high F1-scores SoMer achieves demonstrate the effectiveness of our method.

\begin{table}[ht]
\centering
\caption{F1-Scores on Detecting IO Drivers.}
\resizebox{0.9\linewidth}{!}{
\renewcommand{\arraystretch}{1.2}
\begin{tabular}{p{26mm}p{16mm}p{18mm}p{16mm}}
\Xhline{1pt}
& \textbf{China} & \textbf{Egypt-UAE} & \textbf{Venezuela} \\ \hline
\textbf{BASELINES} & & & \\
Co-Repost & 0.00 & 0.15 & 0.26 \\
Co-URL & 0.19 & 0.27 & 0.30 \\
Hashtag-sequence & 0.08 & 0.20 & 0.05 \\
Text Similarity & 0.13 & \textbf{0.93} & \textbf{0.82} \\ 
SATAR & $0.98 \pm 0.01$ & $0.72 \pm 0.03$ & $0.61 \pm 0.04$ \\
\hline
\textbf{OURS} & & & \\
Temporal & $0.96 \pm 0.01$ & $0.41 \pm 0.13$ & $0.57 \pm 0.06$ \\ 
Textual & $0.97 \pm 0.01$ & $0.69 \pm 0.06$ & $0.67 \pm 0.04$ \\ 
Temporal+Textual & $0.98 \pm 0.01$ & $0.77 \pm 0.05$ & $0.77 \pm 0.04$ \\ 
\textbf{SoMeR} & $\mathbf{0.99 \pm 0.01}$ & $0.85 \pm 0.04$ & $\mathbf{0.82 \pm 0.04}$ \\
\Xhline{1pt}
\end{tabular}}
\label{tab:coord_performance}
\end{table}


\subsubsection{Ablations}
Next, we perform an ablation study on our model to dissect the impact of different features and modules we use. The Temporal model only uses timestamps and three-day post counts as the feature but do not use any textual or network features. The Textual model only uses the \textit{average} textual embedding over three-day intervals, which does not reflect temporal activity. The Temporal+Textual model uses timestamps and the \textit{summed} textual embeddings over three-day intervals, as described in Detecting IO Drivers. This reflects both temporal activity and textual features. The full model SoMeR incorporates network feature on top of the temporal and textual features. All these models include the profile features. We observe that the full model has better performance than any of the ablated models, implying the benefit of including the network link prediction objective. In addition, Temporal+Textual model performs better than Temporal and Textual models alone, indicating that each of the four features plays a role to detect IO drivers. For the China campaign, all ablated and full models have high F1 scores. First, this implies the importance of the temporal features. The Textual model uses \textit{average} textual embeddings over three-day intervals. This lowers but does not completely eliminate temporal factors, and can result in the similar behaviors to the Temporal model. We further confirm the importance of combining different modules by t-SNE of the embeddings generated by different modules in Figure~\ref{fig:coord_tsne} in Appendix. 

The performance of different SoMer modules and different baseline approaches vary across different campaigns. For example, the baseline Text Similarity performs well for Egypt-UAE and Venezuela but poorly for China. On the other side, our models using Temporal features show excellent performance on the China campaign. This shows that campaigns use different tactics and highlights that relying on a single behavior pattern is insufficient, whereas our framework's multi-view approach enables better detection.


\section{Uncovering Shifts in Polarized Discussions} 
\label{section:polarization}
The U.S. society has grown increasingly more  polarize. Not only do liberals and conservatives hold sharply different opinions on a range of issues~\cite{pew2017partisan}, 
but they also have more negative feelings towards members of the other party, compared to members of their own party~\cite{iyengar2018strengthening}. These differences show up not only in political speech, but also in the everyday behaviors~\cite{dellaposta2015liberals}, on on social media platforms, where liberals and  conservatives segregate themselves in different echo chambers, they are reflected in the network structure~\cite{conover2011political}. 
Prior research has found that events can shift public opinion, polarizing the population~\cite{hebbelstrup2023event,jiang2020political,rao2023tracking}.
While many alternative methods for measuring  polarization on social media exist, e.g., using   network analysis~\cite{conover2011political}, interaction behaviors and emotions~\cite{del2016echo}, and community embeddings~\cite{waller2021quantifying}, they are not well suited for measuring shifts in polarization. 
Leveraging our multi-view user representation learning, we provide a new way to measure changes in polarization after significant events by tracking user embeddings, and apply it to measure polarization in the online discussions about abortion, a highly contentious issue in the American society. Our framework allows us to isolate specific topics and identify ones that grew more polarized.


\subsection{Data}
On June 24, 2022, the SCOTUS 
struck down federal protections for abortion rights. This event sparked many online discussions, in which users with different political ideologies expressed distinct views~\cite{rao2023tracking}. We study a public  dataset~\cite{chang2023roeoverturned} containing tweets with abortion-related keywords, such as ``roevswade'', ``prochoice'', and ``prolife''. The data spans the entire year of 2022. We select English posts in the U.S. from users with at least 20 posts. This gives us about 10M posts from 121K users. Using methodology described in \citet{rao2023tracking}, we identify each user's political ideology (liberal/conservative), leaving us with 103K liberals and 18K conservatives. We use the learned user embeddings to track how these two ideological populations change in the user embedding space.

\subsection{Measuring Event-Driven Polarization}
We start with pre-training, which learns a user representation space for all liberal and conservative users using their posting histories in 2022. We perform the same text preprocessing as described in Detecting IO Drivers, except that we use \texttt{sentence-transformers/all-mpnet-base-v2}\footnote{\url{https://huggingface.co/sentence-transformers/all-mpnet-base-v2}} for this English-only dataset. We also use following and follower counts as profile features and use a repost (i.e., retweet) network to train link prediction objective. 

Next, we fine-tune the model via few-shot learning with 358 political elites\footnote{\url{https://github.com/sdmccabe/new-tweetscores}}, such as U.S. politicians, with a known ideology. 
This aligns the embedding space to political ideology. We use a two-layer feed-forward network on top of SoMer and fine-tune the model as described in Detecting IO Drivers.  
Figure~\ref{fig:polarization_tsne} in Appendix shows t-SNE \cite{van2008visualizing} representations of the learned embedding space for all users for the year 2022 both with and without fine-tuning. We see the separation of populations, with conservatives clustered in some regions, whereas liberals, who are 85\% of all data, are distributed across the entire space. The analysis we discuss below describes results using the fine-tuned politically-aware embedding, but we find the same trends using the pre-trained embeddings.

\begin{figure}[ht]
\centering
\includegraphics[width=0.9\linewidth]{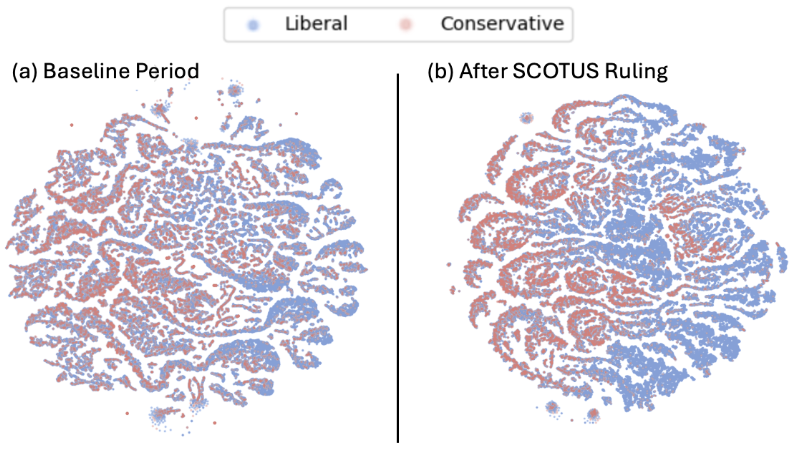}
\caption{Users shifted in the embeddings space after the SCOTUS abortion ruling. Points in (a) are encoded with user post histories between January 1st to May 2nd, 2022. Points in (b) are encoded with the post histories from the same users between June 24th to November 8th, 2022. Points in (a) and (b) are both projected in the same embedding space.}
\label{fig:polarization_shift_tsne}
\end{figure}

In this politically-aware embedding space, to identify shifts in ideological polarization driven by the SCOTUS ruling, we measure how embeddings changed for users before versus after the ruling. We select the period of January 1st to May 2nd 2022 as the baseline period to avoid interference from a leak about this ruling on May 3rd, 2022. We then determine the period to observe the impact of ruling to be June 24th to November 11th 2022. We select this end date to reduce the confounding effect of the 2022 US midterm elections. Next, we select users who posted in both time periods, take their posting history in these two periods separately, and project these users onto the same politically-aware embedding space we have learned. By visualizing the embeddings of the same users in the baseline period and after the SCOTUS ruling, we find a clear shift especially in the conservative population (Figure~\ref{fig:polarization_shift_tsne}). In the baseline period, conservative users were more uniformly distributed in the t-SNE embedding space, but they moved closer together after the SCOTUS ruling. This indicates that the conservative users became more similar in the content of their posts or in their behaviors. 

\begin{figure}[ht]
\centering
\includegraphics[width=1\linewidth]{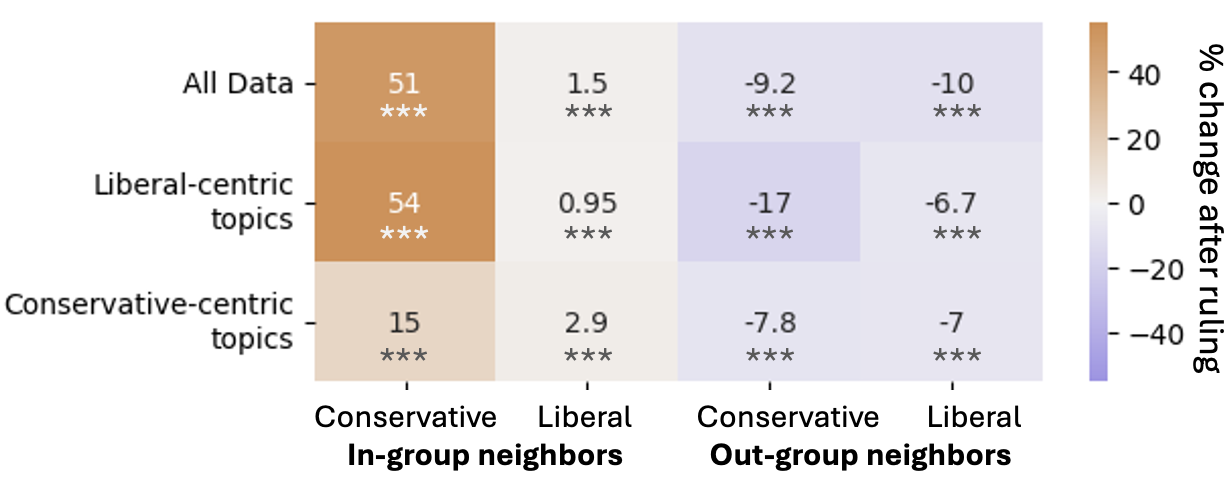}
\caption{Users with same ideology moved closer after SCOTUS abortion ruling, and users with different ideologies moved away. The color represents the \textit{percent change in the mean of these nearest neighbor metrics across populations} from baseline period to after ruling period. *** indicates that the means are significantly different in two time periods with p-value $< 0.0001$.}
\label{fig:polarization_heatmap}
\end{figure}

To further quantify this effect, we find the k-nearest-neighbor (kNN) and check  for each user the percentage of neighbors with the same ideology (in-group) and different ideology (out-group).
We can infer how clustered each population is by the share of the nearest neighbors who are from their in-group. On the other hand, the share of out-group neighbors tells us how far away the two ideological populations are. 
Figure~\ref{fig:polarization_heatmap} shows the percent change in the mean of these four metrics across populations, from the baseline period to the after the ruling. Consistent with Figure~\ref{fig:polarization_shift_tsne}, we see in the ``All Data'' row that the share of in-group neighbors increased for both conservatives and liberals and the share of out-group neighbors decreased. These indicate that \textit{users with same ideology moved closer together, and users with different ideologies moved farther apart in the embedding space}, implying that polarization increased. The conservatives especially became more clustered after the ruling. We perform the same analysis with k=50, 200, 500 nearest neighbors, all showing the same trends.

To dig deeper, we explored how this effect depends on the topic users discuss. \citet{rao2023tracking} identified topics discussed in each post, including religion, bodily autonomy, fetal rights and women's health. We created one subset with posts related to liberal-centric topics, i.e. bodily autonomy and women's health, and another subset with  posts related to conservative-centric topics, like religion and fetal rights. Then we perform the same analysis for each subset. Figure~\ref{fig:polarization_heatmap} shows a consistent overall trend that users with same ideology move closer and users with different ideologies move farther away. Interestingly, we also observe that \textit{each population coalesced when discussing partisan topics promoted by the opposite ideology} - in-group neighbors of conservative users increased more on Liberal-centric topics than on Conservative-centric topics, and similarly in-group neighbors of liberals increased more on Conservative-centric topics than on Liberal-centric topics. Users ``united against a common enemy'' in these online discourses.

\subsection{Ablation}
\begin{figure}[ht]
\centering
\includegraphics[width=1\linewidth]{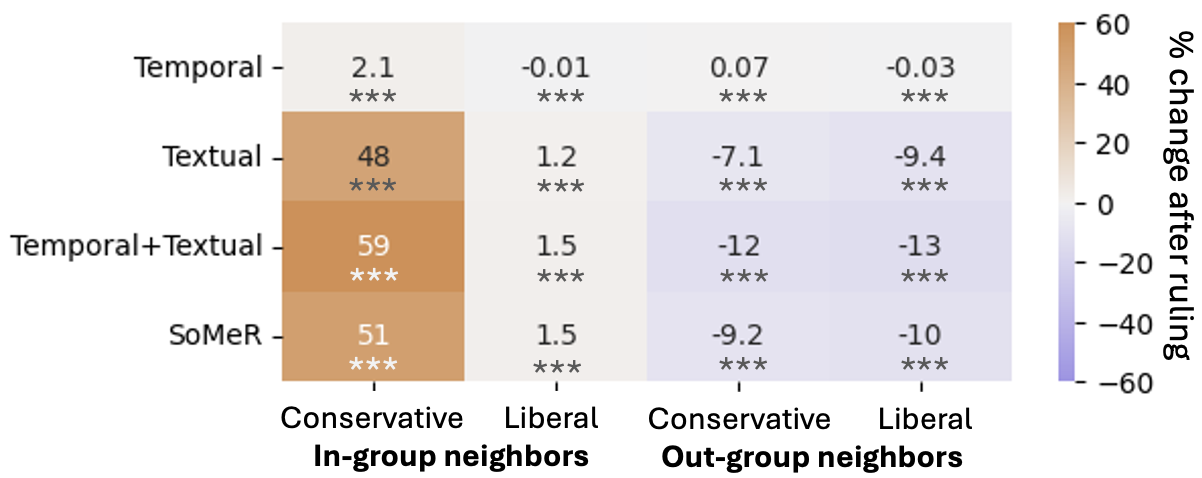}
\caption{Comparison of changes observed in the embedding spaces learned by ablated models and the full model. Temporal features contributed little whereas textual features are the greater factor. The color represents the \textit{percent change in the mean of these four metrics across populations} from baseline period to after ruling period. *** indicates that the means are significantly different in two time periods with p-value $< 0.0001$.}
\label{fig:polarization_ablation}
\end{figure}

We again perform an ablation study to assess how different types of evidence contribute to the shift of user embeddings we observe. Figure~\ref{fig:polarization_ablation} compares three ablated models and the full model. Using the Temporal model results in very little change in all four metrics, indicating that users did not change their activity patterns much after the ruling. Instead, we see much bigger changes when using the Textual model. This implies that users of different ideologies diverged more in the topics and content they discussed after the ruling. Note that the full model SoMeR gives smaller changes than using Temporal+Textual model. During pre-training, the model learns the repost network aggregated over the entire year of 2022, including baseline period and after ruling period. With the repost network not changing between these two periods, user embeddings change less.

\section{Predicting Participation in Hate Subreddits} \label{section:hate_reddit}

Online hate communities have numerous harmful effects on users\cite{schmitz2024users} by radicalizing them \cite{PIRUS} and increasing hate crimes \cite{FBIhate}. While there has been interest in what drives people to hate groups \cite{russo2024stranger}, this analysis was often based on case studies of individual communities. Less-understood is why users may be susceptible to extremism and hate. We aim to use SoMeR to predict whether a Reddit user who had never participated in any hate subreddits will become active in one weeks or even months in the future (on average a user's last post fed into SoMeR was \emph{22 days} before their first post in a hate subreddit). We believe this can help explain what  drives people to join hate groups.



\subsection{Data}

We use a list of 168 hate subreddits collected by \cite{hickey2024peripatetic}, which consists of large banned subreddits with hate speech. We manually inspected the subreddits to verify that they were indeed hate subreddits. We then extract all users who have ever posted in these subreddits until June, 2022 using data stored in Academic Torrent~\footnote{\url{https://files.pushshift.io/reddit/}}. We take a random 200K subsample of all the hate users we collected. To construct a control user group, we collect the user names of 100K random users who ever made a post on Reddit until June, 2022. 

We then collect all posts made by each of these random and hate users, and matched random and hate users such that random users and hate users were active at roughly the same time to make them less trivially distinguishable \cite{luceri2023unmasking}. 
To match random users with hate users, we kept track of which hate users had already been paired. For each random user, we filtered hate users who hadn't been matched, ensuring their first post was before becoming active in a hate subreddit. Then, we randomly selected a hate user active during the same time period as the random user. During training and testing, for each hate user, we only use their posts and comments before the date they became active in a hate subreddit, and we use this same time range for each matched control user.
Within this dataset, we randomly select 4K matched pairs (8K users total) to have a manageable data size for fast experimenting. We keep the overall time range of the dataset to be January 1, 2013 to December 31, 2022, because of the low amount of data before 2013. We then remove low-activity users who have posted and commented less than 10 times. This results in 3771 hateful users and 3444 control users, and 2.3 million posts and comments in total. We split data by 70\%-15\%-15\% for training, validation and testing, respectively.

\subsection{Pre-training and Supervised Classification}
We perform pre-training and supervised fine-tuning to predict whether a user would become active in a hate subreddit later. We first use the same data processing procedure as in~\ref{section:coord_methods}, pre-processing text by removing URLs, emojis and all non-ASCII characters, using sentence-BERT \texttt{all-mpnet-base-v2} to compute the embeddings , and computing the first five components from PCA as the textual features. We aggregate data for each user by summing up their PCA embeddings of their posts in 3-day intervals and taking the middle day in the interval as the corresponding timestamp. This provides us with both the temporal and textual features. 
For the network module, 
a reasonable option we have attempted is to use co-subreddit membership as the linkage (similar users participate in the same subreddits). However, users join subreddits at different times, resulting in different co-activity network from time to time. These co-activity networks for user A may include user B who already joined the hate group, creating information leakage between training and testing. Moreover, because there is no fixed co-activity network, incorporating this feature into the training is very data-intensive and computationally infeasible. For this reason, we had to exclude the co-subreddit membership network feature, but encode users' subreddit membership information as the profile feature, by gathering the 50 most-posted subreddits of each user, and use a lookup-table encoder to generate the profile embeddings.

With all the above features, we then pre-train the model to learn a user embeddings space in which users with similar behaviors are closer together. Next, we perform supervised fine-tuning using a two-layer feed-forward network for binary prediction of whether the user will become active in a hate subreddit later (hyperparameters and training procedure same as in Detecting IO Drivers). 

\subsection{Model Performance and Ablation}

There is no prior task-specific research for predicting user participation in hate communities, hence we first compare SoMeR with a simple baseline---using sentence-BERT embeddings averaged over user posts prior to becoming active in a hate subreddit to train the same two-layer feed-forward network for binary prediction. We also evaluate the multi-view user representation model SATAR as a strong baseline (see SATAR evaluation details in Appendix.) We evaluate all models with randomly bootstrapping training, validation and test data 10 times. Table~\ref{tab:hate_reddit_performance} shows that SoMeR significantly outperforms the BERT baseline by 9\% and SATAR by 20\% on F1-scores, indicating the effectiveness of our method. SATAR has surprisingly lower performance, possibly because its self-supervised objective is not fitted for detecting hate behavior.

\begin{table}[ht]
\centering
\caption{F1 Scores on Predicting User Participation in Hate Subreddits.}
\resizebox{0.7\linewidth}{!}{
\renewcommand{\arraystretch}{1.2}
    \begin{tabular}{lc}
    \Xhline{1pt}
    & \textbf{F1-Scores} \\ \hline
    \textbf{BASELINE} & \\
    BERT & $0.69 \pm 0.01$ \\
    SATAR & $0.59 \pm 0.05$ \\
    \hline
    \textbf{OURS} & \\
    Temporal & $0.74 \pm 0.01$ \\
    Textual & $0.76 \pm 0.01$ \\
    Temporal+Textual & $0.77 \pm 0.00$ \\
    Temporal+Textual+Profile & $\mathbf{0.78 \pm 0.01}$ \\
    \Xhline{1pt}
    \end{tabular}}
    \label{tab:hate_reddit_performance}
\end{table}

Table~\ref{tab:hate_reddit_performance} also shows our ablation study, where we see that the Temporal+Textual model outperforms Temporal only and Textual only models, and using Temporal, Textual and Profile further improves the F1 score. This indicates that all these modules are important. We confirm this with t-SNE plots of embeddings generated by different ablated models in Appendix Figure~\ref{fig:hate_reddit_tsne}.


\section{Conclusion}
In this work, we propose a universal multi-view user representation learning framework \textbf{SoMeR}. Our framework learns from a variety of user features including 1) temporal activities, 2) texts of their posts, 3) profile information and 4) network connections. We show it is versatile and generalizable to different downstream tasks and across different social platforms, including detecting IO drivers, measuring online political polarization, and predicting future user participation in hate subreddits.

There are several limitations and promising avenues for future exploration. First, due to computational constraints, we applied PCA for dimensionality reduction on BERT embeddings when processing user history textual data. We had explored alternative methods like UMAP~\cite{mcinnes2018umap}, but UMAP on big datasets (often 10M or more) requires significant time and compute as well, which is a known challenge. More research is needed to see whether more PCA dimensions, or other dimensional reduction techniques, could improve the performance. 
Second, using the network features has been shown to contribute in model learning. 
Future work may use network graph embedding~\cite{grover2016node2vec} to better learn network-level representations. 
Third, while video, image, and audio modalities have been explored in representation learning, challenges remain in aligning heterogeneous modalities while preserving their characteristics. Pre-training these models often relies heavily on the availability of well-aligned multimodal datasets\cite{xu2023multimodal}. 
Thus, current research on universal URL incorporating these modalities is limited. We acknowledge their potential and aim to incorporate them into future iterations of our framework. 
Fourth, in measuring polarization application, the machine-labeled user political ideologies, although extensively validated by \cite{rao2023tracking}, are not perfect, thus other ideology detection methods should be analyzed to test the robustness of these results. Last but not least, the utilization of temporal features also opens up exciting possibilities to track how users and communities evolve over time. Change point detection on user representations~\cite{guo2024pulse} will be an interesting direction to expand this framework to a more powerful tool.






\bibliography{ref}


\appendix

\section{Ethical Statement}
Our framework utilizes social media data collected from X and Reddit. While the raw data may include personally identifiable information or offensive content, we have taken proactive steps to mitigate privacy and ethical concerns. Specifically, we have documented the data and model usage processes in detail and implemented measures such as removing mentions (@) to minimize the use of personal information. All presented results are aggregated, ensuring no specific account information is disclosed. In addition, we acknowledge the potential risks associated with our framework, including its misuse for harmful purposes, such as predicting future user participation in hate communities. To address these concerns, we have carefully documented the ethical considerations of our framework and committed to tightly regulating training data access, as well as closely monitoring its applications. These measures are intended to mitigate risks to the greatest extent possible while enabling responsible scientific research and ensuring compliance with privacy norms and ethical standards. By emphasizing privacy, fairness, and transparency, we aim to contribute positively to the understanding of social media behaviors while minimizing risks of harm. We remain committed to refining our approach to align with evolving ethical standards and societal expectations.

\section{Appendix}%

\subsection{Verifying Model Validity with Synthetic Data} \label{section:appendix_synthetic}

In this section, we use ablation to test the effectiveness of SoMeR embeddings.

First, we generate two datasets with different number of clusters. These clusters, each containing 1000 samples, mimic user populations with different behaviors. Each sample has a time series with length of 400 in the time domain, and five independently modeled feature categories. We draw each sample and its values of each feature independently from Poisson distributions $X \sim Poisson(\lambda)$ to mimic the daily post counts, and control $\lambda$ for noise level. In addition, we inject peaks into the timeline of these samples, to mimic significant events happening. A peak is independently drawn from a Poisson distribution $X_{peak} \sim Poisson(\lambda_{peak})$ and then \textit{imposed} on top of the baseline.

The first dataset simulates an easy scenario and contains three clusters which are more distinguishable from each other. We make it simple by aligning the events (injected peaks) at the same time points for all the samples in each cluster. 

\paragraph{Data 1} Simple scenario with 3 clusters:
\begin{itemize}
    \item \textit{Cluster 0}: all samples have three peaks at specific time points, each peak lasts for 3 time steps and are drawn from $Poisson(\lambda_{peak}=10)$; samples have higher activity levels - feature values vary by changing their $\lambda \in range(start=4,stop=6,step=0.5)$.
    \item \textit{Cluster 1}: all samples have three peaks exactly same as in Cluster 0; samples have lower activity levels - feature values vary by changing their $\lambda \in range(start=0.5,stop=2,step=0.5)$.
    \item \textit{Cluster 2}: all samples have three peaks different from Cluster 0; samples have lower activity levels - feature values vary by changing their $\lambda \in range(start=0.5,stop=2,step=0.5)$.
\end{itemize}

We run our model with just the textual (represented by these activity feature values) and the temporal modules (represented by the indices of the data points in each sample). The hyperparameters and the setups are the same as described in Methods.

\begin{figure}[h]
    \centering
    \includegraphics[width=1\linewidth]{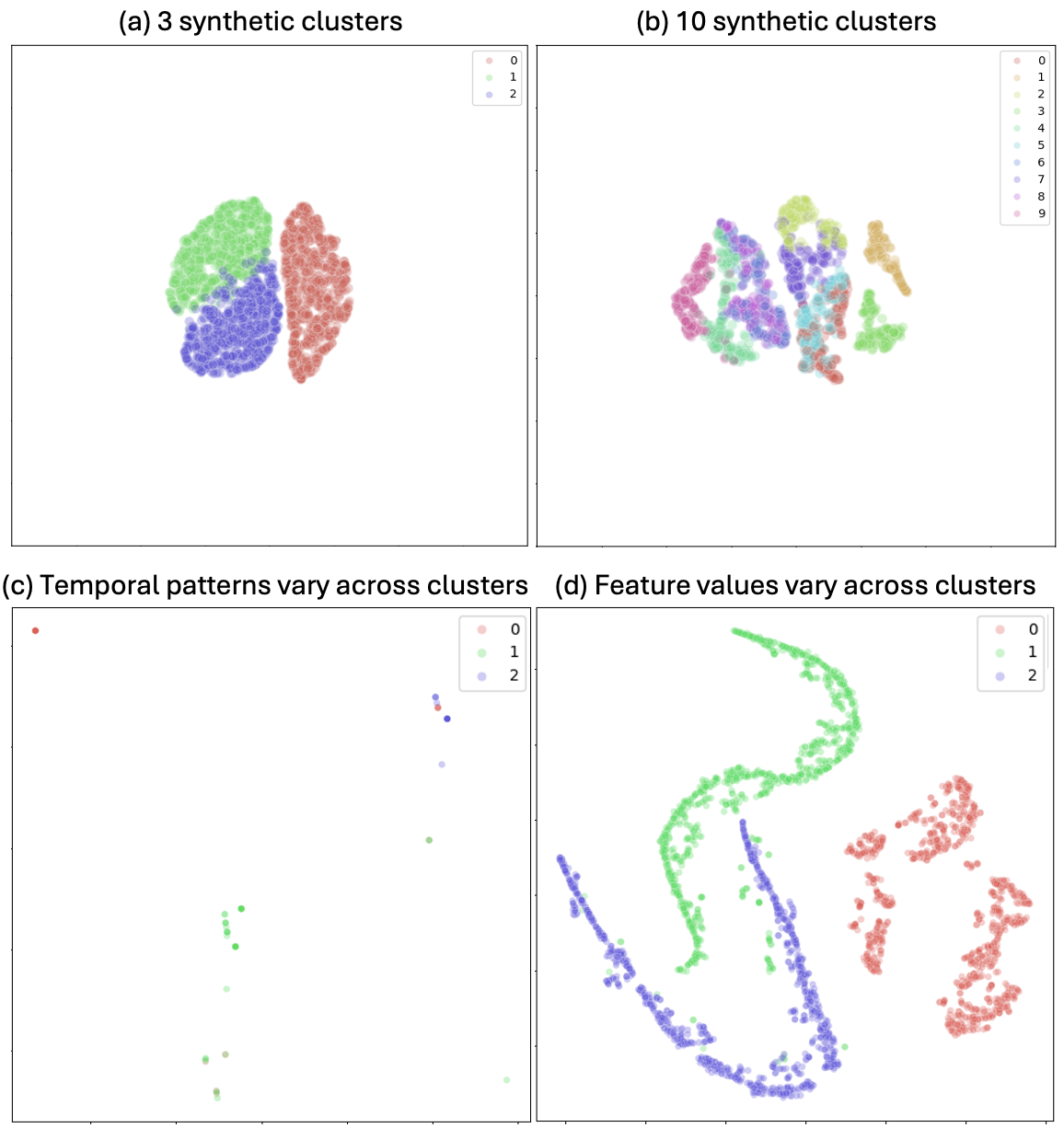}
    \caption{t-SNE of embedding spaces learned from synthetic datasets that (a) simulates a simple scenario with three clusters, (b) simulates a hard scenario with 10 clusters and more noise, (c) simulates the scenario that temporal patterns vary across clusters, and (d) simulates the scenario that feature values across clusters.}
    \label{fig:synthetic_data}
\end{figure}

From Figure~\ref{fig:synthetic_data}(a) we see that our approach nicely distinguish these three clusters. Interestingly, Cluster 1 and 2, having the same population activity level are closer, whereas Cluster 0 with higher activity levels are farther. This implies that activity level, rather than temporal patterns of peaks, might have a heavier impact on the embedding space learned.

The second dataset simulates a harder scenario where the data contains 10 different clusters. Some clusters are more similar with each other,  and some clusters have higher noise level. Thus these clusters are harder to distinguish from each other.

\paragraph{Data 2} Harder scenario with 10 clusters: 
\begin{itemize}
    \item We vary the number of peaks across different clusters: $\#peaks \in [2,3,4,5]$. For clusters with the same number of peaks, the positions of the peaks are the same.
    \item We vary the intensity of peaks across different clusters. The peak intensity is changed with $\lambda_{peak}$ when drawing from the Poisson distribution. $\lambda_{peak} \in [1,2,5,10]$;
    \item We vary the activity levels across different clusters. Five clusters have lower activity level ($\lambda \in range(start=0.5,stop=2,step=0.5)$), and the other five clusters have higher activity level ($\lambda \in range(start=4,stop=6,step=0.5)$).
\end{itemize}

Figure~\ref{fig:synthetic_data}(b) shows the learned embeddings from 10 different clusters. We can see that most clusters are well separated from the others, whereas some clusters that are more similar are closer to each other or overlapping. In general, the samples from each cluster are closely placed in the embedding space, meaning that our model has learned the similarity of samples very well.

Furthermore, to show that it can learn from both temporal activities (timestamps in the triplets) and textual features (features and values in the triplets). We design the following two datasets. One contains three clusters, of which member samples have highly similar temporal patterns but very distinct feature values. In addition, the temporal patterns among the three clusters are different. This is to test if the model can learn heterogeneity in terms of temporal patterns. The second dataset contains three clusters of which member samples have highly similar feature values but distinct temporal patterns. This is to test if model can learn from heterogeneous features and values.

\paragraph{Data 3} Temporal patterns vary across clusters:
\begin{itemize}
    \item \textit{Cluster 0}: all samples  have three peaks at specific time points and each peak lasts for 3 time steps; 
    \item \textit{Cluster 1}: all samples have three peaks different from Cluster 0 and each peak lasts for 3 time steps;
    \item \textit{Cluster 2}: all samples have six peaks at specific time points and each peak lasts for 3 time steps.
\end{itemize}
Within each cluster, samples have low activity level - feature values vary by changing their $\lambda \in range(start=0.5,stop=2,step=0.5)$.

\paragraph{Data 4} Feature values vary across clusters:
\begin{itemize}
    \item \textit{Cluster 1}: all samples have low activity ($\lambda=0.5$)
    \item \textit{Cluster 2}: all samples have medium activity ($\lambda=2$)
    \item \textit{Cluster 3}: samples all have high activity ($\lambda=5$)
\end{itemize}
For all three clusters in Data 4, each sample has three peaks randomly placed in the time domain ($\lambda_{peak}=10$ when drawing for the peak); and each peak lasts for 3 time steps.

Figure~\ref{fig:synthetic_data} shows that on both datasets we can see the segregation of clusters in the t-SNE of embedding space. Surprisingly, on data 3 which varies in temporal patterns across clusters, clusters are very condensed with many samples overlapping on top of each other. For example, almost all the samples in the red cluster in Figure~\ref{fig:synthetic_data}(c) are concentrated on the top left corner. This is probably due to the highly synthetic data we created where all samples have the exact temporal patterns in a cluster. It also indicates that the model can capture the temporal feature well, producing highly similar embeddings for samples with similar temporal patterns. In Figure~\ref{fig:synthetic_data}(d), clusters vary in feature values, specifically the amount of activities. We also observe that data are segregated in alignment with the ground truth cluster labels, indicating the model's ability to learn heterogeneity in features and values. In summary, this part shows that our model can learn from both temporal and feature values in the data.

\subsection{Hyperparameters} \label{section:hyperparameters}

Hyperparameters used in this work and their selection procedure are presented in Table~\ref{tab:hyperparameters}. We used 2 RTX A6000 GPUs for training.

\begin{table*}[]
    \centering
    \resizebox{1\linewidth}{!}{
    \renewcommand{\arraystretch}{1.2}
    \begin{tabular}{l|l|l|l}
    \Xhline{1pt}
         & \textbf{Hyperparameter} & \textbf{Value} & \textbf{Grid Search / Notes} \\ \hline
        P & \begin{tabular}{@{}l@{}}hidden dimension for\\transformer key query and value\end{tabular} & 16 & P = K/H from prior work~\cite{tipirneni2022self} \\ \hline
        K & hidden dimension for other model weights & 64 & [32, 64, 128] \\ \hline
        L & number of transformer layers & 2 & \begin{tabular}{@{}l@{}}small transformer sufficient\\for our data\end{tabular} \\ \hline
        H & number of attention heads & 2 & \begin{tabular}{@{}l@{}}small transformer sufficient\\for our data\end{tabular} \\ \hline
        $\tau$ & temperature scaling parameter & \begin{tabular}{@{}l@{}}0.5 for IO driver detection \\ 3 for measuring political polarization \\ 3 for predicting hateful Reddit users\end{tabular}  & [0.5, 1, 3, 6] \\ \hline
        $\gamma$ & \begin{tabular}{@{}l@{}}noise parameter in data augmentation\\for contrastive learning\end{tabular} & 2 & [0.1, 0.5, 1, 2, 5] \\ \hline
        $\lambda$ & loss balancing parameter & 1 & [0.1, 1, 5, 10] \\ \hline
        lr & learning rate & 5e-5 & \\ \hline
        e & maximum epoch & 60 & early stopping based on val. loss \\ \hline
        b & batch size & \begin{tabular}{@{}l@{}}32 when link prediction module used\\128 for other cases\end{tabular} & \begin{tabular}{@{}l@{}}[32, 64, 128]\\doesn't impact performance\end{tabular} \\ \hline
        dr & drop-out rate in supervised finetuning & 0.3 & [0, 0.3, 0.5] \\
        \Xhline{1pt}
    \end{tabular}}
    \caption{Hyperparameters used in this work.}
    \label{tab:hyperparameters}
\end{table*}

\subsection{Evaluating SATAR Baseline} \label{section:appendix_satar}

\subsubsection{Detecting IO Drivers}
We compare our framework with a strong baseline SATAR~\cite{feng2021satar} on detecting IO drivers. SATAR is a more advanced multi-view user representation learning approach that combines textual, profile and network features. The authors uses it for social bot detection. To adapt SATAR for IO driver detection, we keep the model architecture and most of the setup the same as the original paper~\footnote{Code for SATAR is publically available at \url{https://github.com/BunsenFeng/SATAR/tree/main}}. For SATAR's profile property module, it originally uses 15 true-or-false property items, e.g. ``profile uses background image'',  5 numerical property items, e.g. ``favorites count'', and ``location''. However, the authors did not specify the 15 binary attributes. In addition, some Twitter user attributes like location are inaccessible in our data. We therefore use the textual \texttt{user profile description} as the profile features, encoding the first 10 words into word2vec embeddings. The description often include a short summary of the user and many times include locations too, and we believe it is a good substitute for the profile property features. Similar to the original paper, we first perform self-supervised pre-training for follower count prediction, and fine-tune SATAR with IO driver labeled data. We pre-train for 60 epochs to make sure of model convergence. The following-follower network module in SATAR iteratively uses and updates the neighbor embeddings for each user in the network, resulting in more epochs to converge. We then fine-tune for 5 epochs. The batch size is 4. The data used and the train-validation-test splits are the same as those for training SoMer. Other model structure parameters are kept the same as the original SATAR paper.

\subsubsection{Predicting Participation in Hate Subreddits}
SATAR is originally designed for X platform with many X-specific features. For predicting participation in hate subreddit, we had to adapt SATAR to Reddit by (1) disregard the followinng-follower network feature, (2) using top 50 subreddit membership as the profile feature, and (3) train SATAR with the self-supervised objective to predict total number of ``ups'' a user receives instead of the follower counts. We then supervisedly fine-tune SATAR with hate user labels. These modifications align with our setup for SoMer, offering a fair comparison. We pre-train for 5 epochs. We do not include the following-follower network module for SATAR in this application and therefore the model quickly overfits to the pre-training ``ups'' count signals. We then fine-tune for 5 epochs. The batch size is 4. Other model structure parameters are kept the same as the original SATAR paper.

\subsection{Embeddings for IO drivers} \label{section:appendix_coord}

\begin{figure}[h]
   \centering
   \includegraphics[width=1\linewidth]{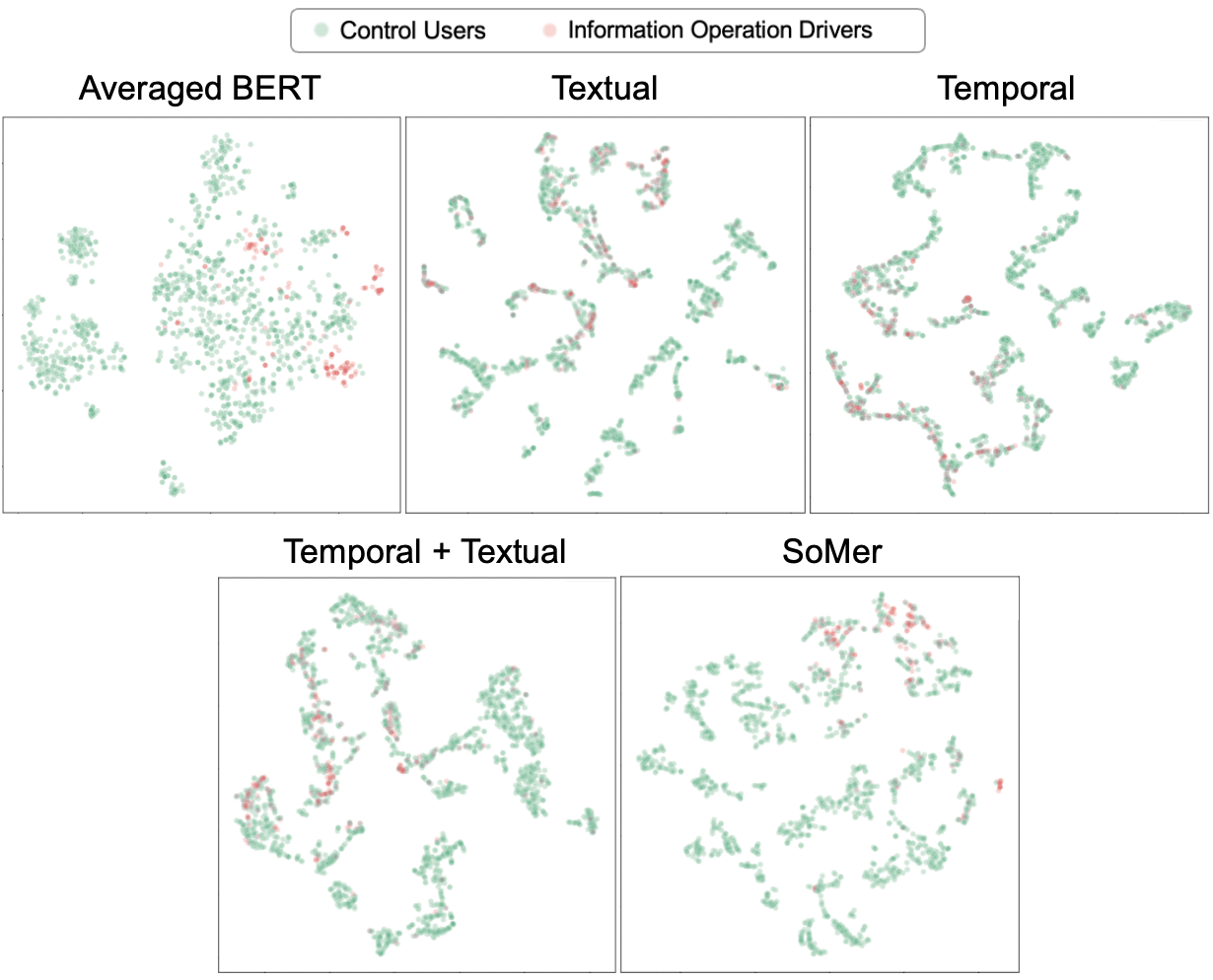}
   \caption{t-SNE of user embeddings in the Egypt-UAE data using different models.}
\label{fig:coord_tsne}
\end{figure}

In the application of detecting IO drivers, to further understand why SoMer performs less satisfactory on the Egypt-UAE data, we investigate in how embeddings look like using different modules and plot the t-SNE in Figure~\ref{fig:coord_tsne}. The sentence-BERT embeddings averaged across the entire time (Averaged BERT in the figure) corresponds to the features used in the Text Similarity baseline model. We notice that the IO drivers already separate well from the control users in this embedding space. This indicates that text similarity indeed is a very important feature in this campaign, consistent with the very high F1 score of this baseline model. The campaign might have promoted to use similar language as their tactic. On the other hand, the embeddings from our approach only using textual features are not separated so well. This again could be due to our dimension reduction step on the sentence-BERT embeddings. Second, Egypt-UAE is the smallest campaign here and our models are pretrained on only 2.5K users. We hypothesize that the models trained on small datasets and under-fitted can also result in less ideal performance. 

Nevertheless, we observe that the full SoMer model can generate more separable embeddings compared to the other ablated models. This is especially obvious when comparing embeddings from Temporal+Textual and the full SoMer model (Temporal+Textual+Network). This shows the significance of the network module, which aligns well with the significant jump of F1-scores between these two models. From these embedding plots, we can see how combining different modules improves model's representation learning ability.

\subsection{Effectiveness of fine-tuning models on partisanship} \label{section:appendix_polarization}
\begin{figure}[ht]
\centering
\includegraphics[width=0.9\linewidth]{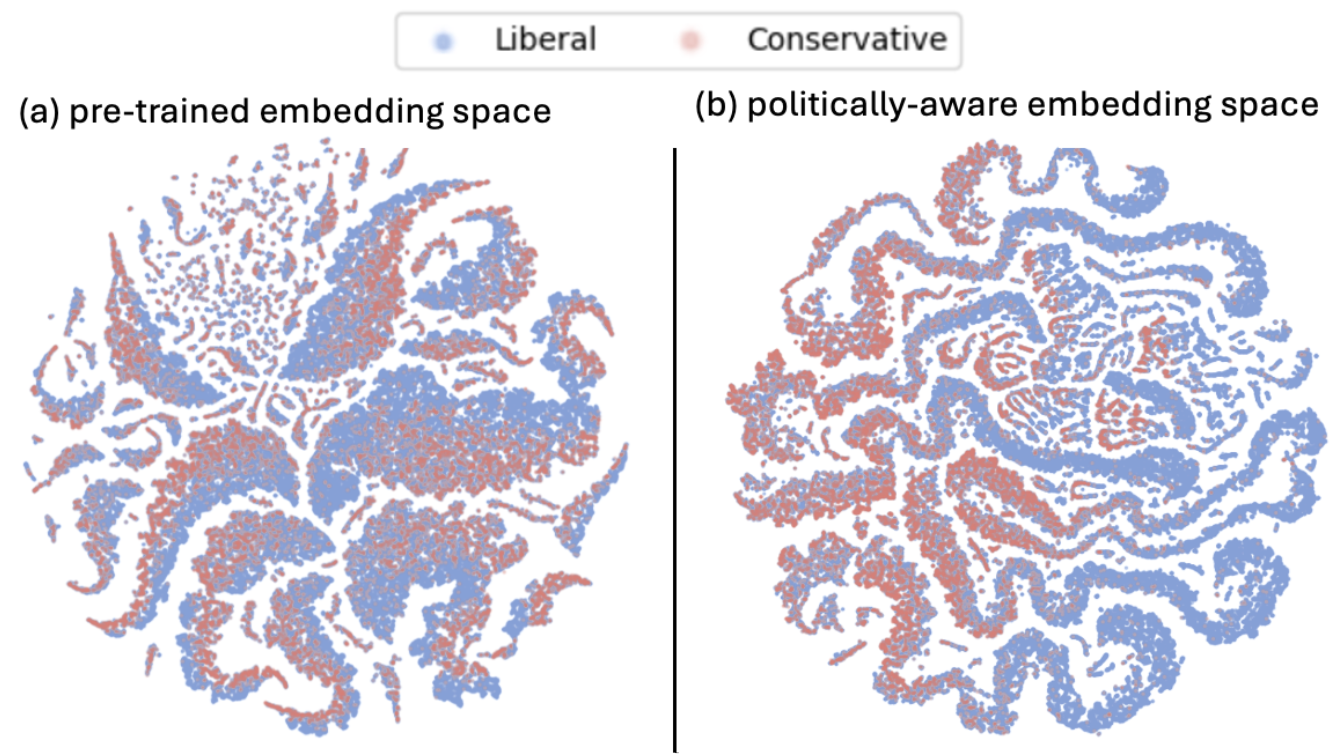}
\caption{(a) t-SNE of pre-trained user embedding space learned with both ideology populations over the entire year of 2022. (b) t-SNE of the space fine-tuned with data from elite users with known ideologies, making the embedding space more politically-aware.}
\label{fig:polarization_tsne}
\end{figure}

In the applications of uncovering polarization, in Figure~\ref{fig:polarization_tsne} we show t-SNE representations of models trained on the SCOTUS decision dataset both with and without fine-tuning. We find that fine-tuning more effectively separates the political ideology dimension along the x-axis.

\subsection{Embeddings for hateful Reddit Users} \label{section:appendix_hate_reddit}

\begin{figure}[h]
   \centering
   \includegraphics[width=1\linewidth]{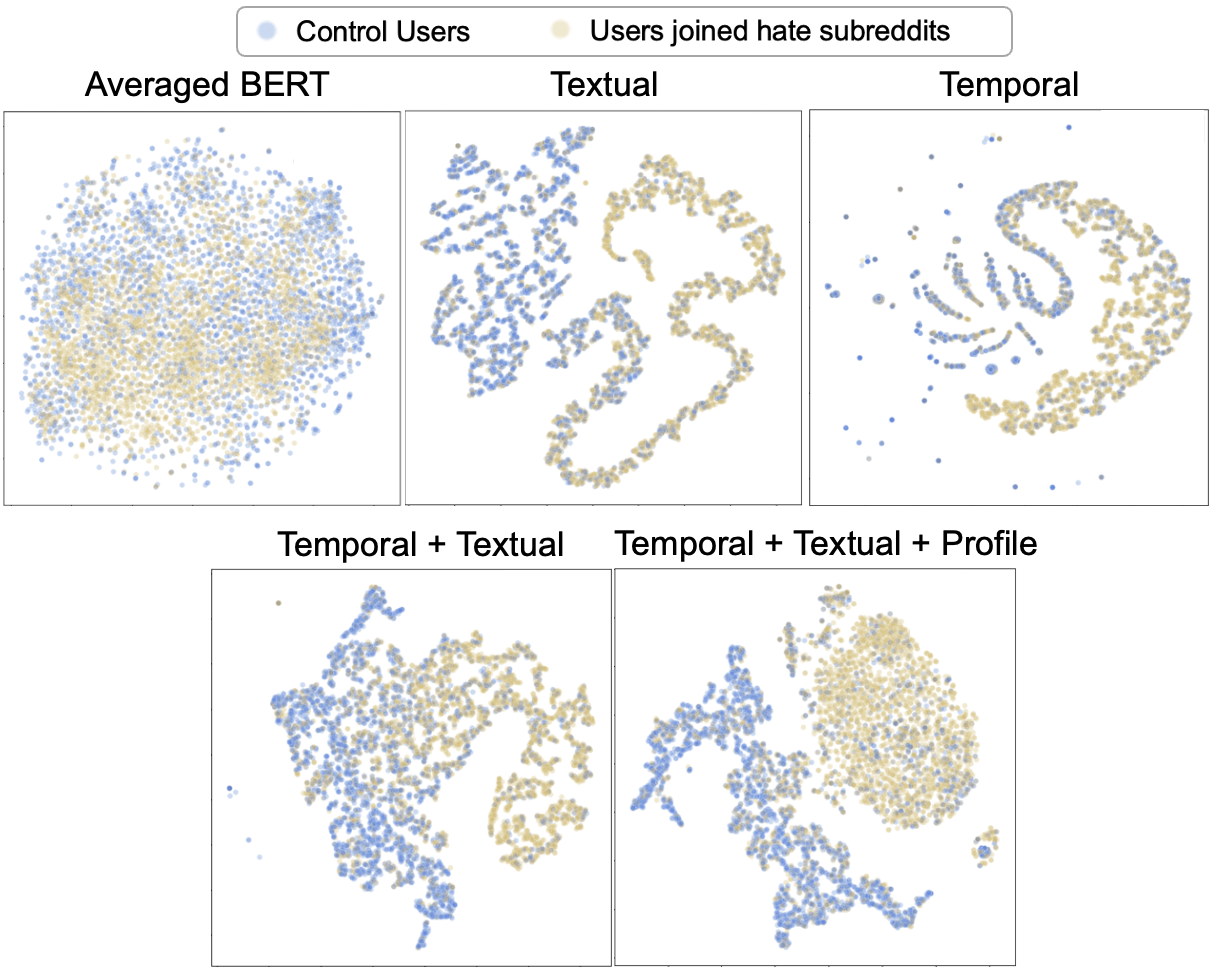}
   \caption{t-SNE plots of embeddings generated by different models.}
   \label{fig:hate_reddit_tsne}
\end{figure}

For predicting hate users in the Reddit data, Figure~\ref{fig:hate_reddit_tsne} shows the t-SNE plots of embeddings learned by different models. We observe that the baseline, sentence-BERT embeddings averaged across time, do not distinguish between control and hateful users, resulting in a lower F1 score on user classification task. It is also interesting to see how Temporal+Textual+Profile model better separates the control and hateful users into two clusters, compared to other ablated models. This demonstrates the benefits of our multi-view approach that combines different features.

\end{document}